\input phyzzx
\def\e{\adveq\eqno{\rm (\chapterlabel\the\equanumber)}}

\def\adveq{\global\advance\equanumber by 1}
\def\myeq{{\rm \chapterlabel\the\equanumber}}
\def\rarrow{\rightarrow}

\def\semidirect{\mathrel{\raise0.04cm\hbox{${\scriptscriptstyle |\!}$
\hskip-0.175cm}\times}}

\def\mod{\mathop{\rm mod}\nolimits}

\def\ref#1{$^{[#1]}$}

\def\r#1{$[\rm#1]$} 
\def\twidle{\tilde}

\def\osum{\mathop\oplus\limits}

\def\e{\adveq\eqno{\rm (\chapterlabel\the\equanumber)}}

\def\adveq{\global\advance\equanumber by 1}
\def\myeq{{\rm \chapterlabel\the\equanumber}}
\def\rarrow{\rightarrow}

\def\semidirect{\mathrel{\raise0.04cm\hbox{${\scriptscriptstyle |\!}$
\hskip-0.175cm}\times}}

\def\mod{\mathop{\rm mod}\nolimits}

\def\ref#1{$^{[#1]}$}

\def\r#1{$[\rm#1]$} 
\def\twidle{\tilde}

\def\osum{\mathop\oplus\limits}

\def\half{{1\over2}}
%


\overfullrule=0pt 
\date{November, 1998} 
\titlepage 
\title{On Conformal Field Theories at Fractional Levels} 
\author{Ernest Baver, Doron Gepner and Umut G\"ursoy}
\address{Department of Physics\break
         Weizmann Institute of Science\break
         Rehovot, Israel\break}
\vskip15pt 
\abstract 
For each lattice one can define a free boson theory propagating on the 
corresponding torus. We give an alternative definition where one employs 
any automorphism of the group $M^*/M$. This gives a wealth of conformal 
data, which we realize
as some bosonic theory, in all the `regular' cases. We discuss the generalization to affine theories.
As a byproduct, we compute the gauss sum for any lattice and any diagonal automorphism.
\endpage

Free bosons systems in two dimensions have been a source of a simple environment in which both string theories and condensed matter systems could be studied. Ideas such as Narain compactification 
\REF\Narain{K.S. Narain, Phys. Lett. B 169 (1986) 41}\r\Narain\
and Orbifolds 
\REF\Orbifold{L. Dixon, J. Harvey, C. Vafa and E. Witten, Nucl. Phys. B 274 (1986) 285}\r\Orbifold\
have shown that although simple, these systems offer a vast richness.
 
In ref. \REF\long{D. Gepner, Caltech preprint CALT-68-1825, Nov. 1992, hepth 9211100}\r\long\ one of this authors described the notion of a  ``pseudo" bosonic system. This is a conformal field theory in which one formally replaces the level in the modular matrix $S$ with a fractional number. We denote this system  
as $M_{1/q}$ where $M$ is the lattice defining the boundary condition of the boson and $q$ is some number strange to 
$|M^*/M|$, where $M^*$ is the dual lattice of $M$. The primary fields are elements of $M^*/M$ and the modular matrix is,
$$S_{\lambda,\mu}=|M^*/M|^{-\half}\exp(-2\pi i q\lambda\mu).\e$$
These bosonic systems are important in other constructions described in ref. \r\long\ of new RCFT, as well as being mathematically intriguing. Actually, the definition of $M_{1/q}$ can be generalized vastly. Consider the group $G\approx M^*/M$. Then for each automorphism of $G$, denoted by $h$, we can define a set of conformal data. The $S$ matrix is given by,
$$S_{\lambda,\mu}=|M^*/M|^{-\half}\exp(-2\pi i h(\lambda)\cdot\mu).\e$$
This $S$ matrix is symmetric, it has the same fusion rules for all $h$ and it generalizes $M_{1/q}$ when one takes
$h(\lambda)=q\lambda\mod M$. For cyclic groups, all automorphisms are of this form. Otherwise, $M_h$ is more general.

Note, in passing that it is not important how we define the scalar product. Take $g(\lambda,\mu)$ to be a real valued, non degenerate, bi-linear, symmetric function on $G\times G$. Then it is easy to show that $g$ can be written as $g(\lambda,\mu)=h^\prime(\lambda)\cdot\mu$ where $h^\prime$ is some automorphism of $G$, and vice versa.
Thus, we can adopt any definition for the scalar product.
 
Now according to the basic theorem of Abelian groups, $G$ can be decomposed as a direct sum of cyclic groups,
$$G\approx \osum_i Z_{n_i}.\e$$
Denote by $\pi_r(\lambda)$ the projection onto the $r$'th space. The most general automorphism, $h$, can then be written as,
$$h(\lambda)=\sum_i q_i \pi_i(\lambda),\e$$
where $q_i$ is any set of integers strange to $n_i$
 We can then write eq. (2) as,
$$S_{\lambda,\mu}=|M^*/M|^{-\half} \exp(-2\pi i\sum_n \pi_n(\lambda)\cdot \pi_n(\mu)q_n).\e$$
Considered then as a direct sum group, all the formulas of ref. \r\long\ apply to each summand, e.g., the modular matrix $T$ and the central charge $c$,
$$T_{\lambda,\lambda}=\exp\left(\sum_i \pi i\pi_n(\lambda)^2 q_n\right),\e$$ 
$$\exp(\pi i c/4)=|M^*/M|^{-\half}\sum_{\lambda\in M*/M}
\exp(\pi i \sum_n q_n \pi_n(\lambda)^2).\e$$
The central charge, $c$, is always an integer (defined modulo 
$8$).
 
It was conjectured in ref. \r\long\ that all such pseudo bosonic systems are in fact isomorphic to some other real bosonic systems (i.e., $q=1$). Our aim here is to prove this conjecture and to describe explicitly the relation.
Several previous examples were known, e.g., $SU(2)_{-1}\approx (E_7)_1$, $SU(3)_{-1}\approx (E_6)_1$.
To put it precisely, $M_h\approx {\twidle M}_1$ as RCFT's if and only if there exist an isomorphism $\phi$,
\def\Mt{{\twidle M}}
$$\phi:M^*/M\rarrow \Mt^*/\Mt,\e$$
in such a way that 
$$\phi(\lambda)^2=\sum_n q_n\pi_n(\lambda)^2 \mod 2.\e$$
This condition ensures that the RCFT $M_1$ will indeed describe the correct conformal data. 
 
The proof of the aforementioned conjecture proceeds in two stages. First we will show that for any lattice the problem amounts to solving it for $SU(p^n)_q$ where $p$ is some prime. Then we will solve it for the latter theories explicitly.
 
Consider then the Abelian group $G\approx M^*/M$. The group algebra of $G$ is the so called fusion algebra (or more precisely a ring over $Z$). $G$ is a finite group. Now, the basic theorem of Abelian groups asserts that $G$ is isomorphic to the group,
$$G\approx Z_{p_1^{n_1}}\oplus\ldots \oplus Z_{p_m^{n_m}},\e$$
where $p_r$ are some prime numbers, $n_r$ are some integers and $Z_k$ is a cyclic group of order $k$. 
  
Inspect the theory 
$$N\approx SU(p_1^{n_1}) \times \ldots \times SU(p_m^{n_m}).\e$$
This theory has exactly the same group $G$ for the fusion rules.

Now a celebrated relation by Verlinde \REF\Verlinde{E. Verlinde, Nucl. Phys. B 300 (1986) 285}\r\Verlinde\ asserts that the modular matrix is the solution of the fusion rule equation,
$$f_{\lambda,\sigma}^\nu={\sum_a S_{\lambda,a} S_{\sigma,a} S^\dagger_{\nu,a} \over S_{0,a}}.\e$$  
Moreover, all the theories $M_h$ have the same fusion rules, for all $h$. It follows that the modular matrices of $M_{h}$ and $N_{h^\prime}$ are the same, with some relation between $h$ and $h^\prime$, for all $h$, since they are solutions to the same equation, and since all the solutions to of this equation are of the form eq. (2). To see that these are all the solutions, with the physical demand of $S_{0,a}>0$, consider the points of the fusion variety (defined by the fusion ring), equal to
$$x_\lambda^\mu=S_{\lambda,\mu}/S_{0,\mu},\e$$
From this we find $S_{\lambda,\mu}$ uniquely, up to some permutation of the rows. By the symmetry of the $S$ matrix,
and the fact that the $x$'s represent the fusion ring, it follows that this permutation must be an automorphism as prescribed in eq. (2).
 
It follows that we need to solve this conjecture only for $SU(m)_{h}$, where $h$ is any automorphism of $G$, and $m$ is a prime power. This is described in the appendix. 

Recall, first an argument for ref. \r\long\ about $h$. Taking 
$$q_i\rarrow s_i^2 q_i,\e$$ 
where $s_i$ is any integer strange to $n_i$, leaves the lattice theory unchanged, since it amounts to the trivial re-labeling of the lattice elements by,
$$\pi_r(\lambda)\rarrow s_r \pi_r(\lambda).\e$$
Thus, we need to consider only one member $h$ of each class modulo a square.

Actually, we assumed in defining the automorphism that the group $G$, $G\approx \oplus_i Z_{n_i}$ is such that $\gcd(n_i,n_j)=1$ for all $i\neq j$. This case is termed `regular'. 
When this is not so,
say $n_i=p^a$, $n_j=p^b$ for some $i$ and $j$, it is possible 
to mix them in the automorphism. We can of course assume that all $n_i=p^{a_i}$ for some $a_i$. The general case, as discussed above is just a direct sum of these. 

Now, let $M$ be any symmetric matrix of integers such that $\det M\neq 0\mod p$. For any such matrix, we can define an automorphism $h$ defined on the basis $\lambda_i$ of $G$,
$$h(\sum_i n_i \lambda_i)=\sum _j M_{ij} n_i \lambda_j.\e$$
Evidently, $h$ is an automorphism, since $M$ is invertible.
We also need to assume,
$$M_{ij} p^{a_i}=0\mod p^{a_j},\e$$
for all $i,j$, which is necessary and sufficient to ensure that $h$ is onto.
Further, we can define a bilinear, symmetric, non--degenerate  
form on $G\times G$ by,
$$K_h (\lambda,\mu)=h(\lambda)\cdot \mu,\e$$
where `$\cdot$' is some standard inner product, say $\sum_n \pi_n(\lambda)\pi_n(\mu)$.
By the arguments above it follows that such forms are in $1-1$ correspondence with the group of such matrices, denoted by $\cal M$.
 
We define a modular matrix, $S$, for any automorphism $h$, by
$$S_{\lambda,\mu}=|M^*/M|^{-\half}\exp(-2\pi i h(\lambda)\mu),\e$$
for the primary fields $\lambda,\mu\in M^*/M$. 
The modular matrix is symmetric (by the symmetry of $h$), and
obeys $S^2=C=\delta_{\lambda+\mu}$, $S^\dagger S=1$.
In fact, $S$ is just a permutation of the $h=1$ matrix, by the permutation $S_{\lambda,\mu}\rarrow S_{\lambda,h(\mu)}$.
This shows that $S$ gives the same fusion rules for all the automorphisms $h$. Moreover, by the arguments above, all the solutions of Verlinde equation correspond to some automorphism. We denote the theory with the $S$ matrix eq. (19) by $M_h$. The dimensions $\Delta_\lambda$ and the central charge $c$ are found to be,
$$\Delta_\lambda=\half h(\lambda)\cdot\lambda \quad \mod 1,\e$$    
$$e^{\pi i c/4}=|M^*/M|^{-\half} \sum_{\lambda\in M^*/M}
e^{\pi i\lambda\cdot h(\lambda)},\e$$
$$T_{\lambda,\lambda}=e^{2\pi i(\Delta_\lambda-c/24)}.\e$$
Curiously, for any lattice $M$ and any automorphism $h$ of $M^*/M$, we find that $c$ is an iteger, suggesting that it is a free boson theory. We have, $(ST)^3=S^2=\delta(\lambda+\mu)$.
 
In passing, we note that one can evaluate the Gauss sum eq. (21) as follows. Let 
\def\hpr{h^\prime}
$\hpr$ be a lifting of the automorphism $h$ to all of $M^*$.
This can always be done in infinite many ways. Assume also
that $\hpr$ has only positive eigenvalues. Then, we define the set of theta functions,
$$\theta_\lambda(\tau,z,u)=e^{2\pi i u}\sum_{\mu\in M+\lambda} e^{\pi i \tau \mu\hpr(\mu)+2\pi iz \hpr(\mu)},\e$$
where $\lambda\in M^*/M$.
Under the modular transformation $S$ we have the metaplectic action,
$$\theta_\lambda(-1/\tau,z/\tau,u-z^2/2\tau)=
(-i\tau)^{-l/2} \sum_\mu S_{\lambda,\mu} \theta_\mu(\tau,z,u),\e$$
where $l={\rm rank}(M)$ and $S$ is given by eq. (19).
Under $T$ we have,
$$\theta_\lambda(\tau+1,z,u)=e^{2\pi i c/24}T_{\lambda,\lambda}\theta_\lambda(\tau,z,u),\e$$
where $T$ is given by eq. (22).
Define the congruence modular subgroup $A_k$ as
$$A_k=\left\{\pmatrix{a&b\cr c&d\cr}\big| a-1=d-1=b=c=0\mod k\right\}.\e$$
Then the theta functions defined above are invariant under the action of $A_k$ (defined by $\tau\rarrow (a\tau+b)/(c\tau+d)$), where $k$ is
$$k=2|M^*/M|.\e$$
Now, $\theta_\lambda(\tau,0,0)$ transform under $S$ 
and $T$ exactly by the matrices eqs. (19,22) (up to a phase). 
We can use them to define the so called cusp functions of the modular group $A_k$, and by that express the real characters in terms of these. This is is under investigation. A byproduct of this is the expression of the theta functions in terms of Dedekind $\eta$ functions,
which is another way to describe the cusp functions. It follows that we can compute 
$$e^{\pi i c/4}=C \sum_\lambda\theta_\lambda(1+i\epsilon,0,0),\e$$
where $C$ is a positive constant,
from the multiplier
system of the $\eta$ function, which is known. This gives a formula for $c$ in terms of the so called Dedekin's symbols.
It follows also that the characters of the RCFT $M_h$ can be expressed in terms of these. Below, we give another expression for the gauss sums.   

Now, not all the elements of $\cal M$ give rise to distinct physical theories. Suppose that
$$M_{h_1}=B^t M_{h_2} B,\e$$
where $B$ is any integer matrix. Then, we can relabel the primary fields, $M^*/M$ by
$$ x\rarrow Bx,\e$$
giving of course the same physical theory. However, for the scalar products we have,
$$K_{h_1}(\lambda,\mu)=K_{h_2}(B\lambda,B\mu),\e$$
showing that $M_{h_1}$ is the same theory as $M_{h_2}$.
This simplifies the problem of classifying the different theories. Define the equivalence relation on $\cal M$ by
$$M\approx B^T M B,\e$$
for any matrix $B$. This is evidently an equivalence relation, and it follows that the physical theories $M_h$ are in correspondence with the quotient set ${\cal N}={\cal M}/\approx$. 
  
Our aim is to find for each lattice $M$ and each automorphism $h\in{\cal N}$ a realization as usual free bosons theory. This is equivalent to finding a lattice $\Mt$ and an isomorphism $\phi:M\rarrow \Mt$ such that
$$\phi(\lambda)^2=\lambda\cdot h(\lambda)\quad \mod 2.\e$$
For an automorphism, $h$, which is not regular, this is still under investigation.
Examples:

1) Consider the lattice $M=SU(3)_1\times SU(3)_1$. Here, $M^*/M\approx Z_3\oplus Z_3$ and we define the automorphism which exchanges the two algebras,
$$h(\lambda_1,\lambda_2)=(\lambda_2,\lambda_1).\e$$
Using eq. (21), we find 
$$e^{\pi i c/4}={1\over3} \sum_{m_1,m_2\mod 3} e^{\pi i \min(m_1,m_2) (3-\max(m_1,m_2))/3}=1.\e$$
Thus $c=0\mod 8$.  

2) Take $M=SU(2)_2\times SU(2)_2$. Assume
$$h(\lambda_1,\lambda_2)=M\pmatrix{\lambda_1\cr\lambda_2\cr},
\qquad {\rm where\ } M=\pmatrix{1&2\cr2&1\cr}.\e$$
We find then
$$e^{\pi i c}={1\over4}\sum _{m_1,m_2\mod 4} e^{{\pi i\over 4}(m_1^2+m_2^2+4m_1m_2)}=e^{-\pi i/2}\e$$
and $c=-2\mod 8$. 

3) Take $M=SU(N)_k$. Then $M$ is generated by the simple roots $\{\alpha_i\}$ and $M^*$ by the fundamental weights
$\{\lambda_i\}$. We have the relations $Nk\lambda_i=0\mod M$
and $k(\lambda_i-i\lambda_1)=0\mod M$. Thus $G=M^*/M$ is the group
$$G\approx {Z_{Nk}^{N-1}\over
Z_N^{N-2}}.\e$$

Consider now the related case of affine theories. for the group $G_k$ with the root lattice $M$, these are defined on the lattice $M\sqrt{k+g}$ ``modulo" the action
of the Weyl group $W$. We may thus use the above results to define new affine theories. 
Take for the modular matrix $S$,
$$S_{\lambda,\mu}=i^{|\Delta_+|} (-1)^s |M^*/(k+g)M|^{-\half} \sum_{w\in W} e^{-
2\pi i w(\lambda+\rho) h(\mu+\rho)/(k+g)},\e$$
where $|\Delta_+|$ is the number of positive roots,
$\lambda,\mu$ are some primary fields, $\rho$ is half the sum of positive roots, and $g$ is the dual Coxeter number. $h$ is any automorphism of the group $G=M^*/(k+g)M$.
(This is defined similarly for a direct sum of groups.) Define, in passing, $\alpha$ modulo $8$ 
$$i^{\alpha/2}=|M^*/M|^{-\half} \sum_{w\in W} (-1)^w e^{2\pi i w(\rho)\cdot h(\rho)/g},\e$$
which is equal, evidently, to
$$i^{\alpha/2}=i^{|\Delta_+|} (-1)^{\bar w},\e$$
where $\bar w$ is a Weyl element defined by
$$h(\rho)=\bar w(\rho) \mod gM.\e$$
The proof is obvious by plugging eq. (42) into eq. (40).
$s\mod2$ is defined by
$$h(\lambda)=w(\rho) \mod (k+g)M,\e$$
where $w\in W$ and
$$(-1)^s=(-1)^w.\e$$
Using the denominator identity of affine algebras \REF\Kac{
V. Kac, Infinite dimensional Lie algebras, Cambridge University press, (1990)}\r\Kac, it can be shown,
$$i^{\alpha/2}=|M^*/gM|^{-\half} \prod_{\alpha>0} 2i\sin({\pi h(\rho)\alpha\over g}),\e$$
where the product is over all the positive roots.
This sign, $s$, was chosen by the physical requirement that the identity elements would have a positive modular elements. 
The automorphism 
$$h(\lambda_i)=M_{ij} \lambda_j,\e$$
must be symmetric, as before, $M^t=M$, along with the condition that
for all $h$ and $w\in W$ exist $w^\prime\in W$ such that
$$hw=w^\prime h.\e$$

As before, $h$ is defined modulo a similarity transformation,
$h\approx B^t h B$.
For each such automorphism $h$, the matrix $S$ is symmetric
and obeys the same fusion rules as the usual one ($h=1$). The case $h(\lambda)=p\lambda$ where $\gcd(p,|M^*/(k+g)M|)=1$
was described in ref. \r\long. In an abuse of notation, we use $h$ to designate $\bar h$ the extentions of $h$ to all of $M$,
$\bar h:M\rarrow M$ and $h(m)=\bar h(m) \mod (k+g)M$.
 
Thus we get many new symmetric modular matrices. However, we need to find a diagonal matrix $T$, such that $(ST)^3=C=\delta_{\lambda,\bar\mu}$. The only obvious candidate for the conformal dimensions is
$$\Delta_\lambda={[(\lambda+\rho) h(\lambda+\rho)-\rho h(\rho)]\over 2(k+g)},\e$$
$$e^{\pi i \Theta/4}=|M^*/(k+g)M|^{-\half} \sum_{\lambda\in M^*/(k+g)M}
e^{\pi i \lambda\cdot h(\lambda)/(k+g)},\e$$
$$T_{\lambda,\lambda}=e^{2\pi i(\Delta_\lambda-c/24)},\e$$
and
$$c=-{12\rho h(\rho)\over k+g}+2|\Delta_+|+4s+\Theta={12\rho h(\rho) k\over g(k+g)}+4r \mod 8,\e$$
where $(-1)^r$ is a sign we were not able to determine (see below).
As a byproduct for the formula for the central charge, eq. (51), which will be proven below, we compute the gauss sum, for any $M$, $k$ and $h$,
$$\Theta={12\rho h(\rho)\over g}-2|\Delta_+|+4 s+4r\mod 8.\e$$
The above generalizes for a multiple of groups (where we define the lattice $M=\oplus_i M_i \sqrt{k_i+g_i}$ and omit the factors of $k+g$).
However, for the definition eq. (48) to make sense we need a stronger condition on $h$,
$$hw=wh,\e$$
for all $w\in W$. This condition, we think, is obeyed only by the product form described in ref. \r\long, $h(\lambda)=p\lambda$, and we find no new solutions for the conformal data, other then the ones denoted $G_{(k+g)/p}^{\rm affine}$. 

We can define the Weyl anti-invariant theta functions by
$$A_{\lambda,k}(\tau,z,u)=\sum_{w\in W} (-1)^w \theta_{w(\lambda+\rho),k+g}^{M,h}(\tau,z,u).\e$$
The $A_{\lambda,k}(\tau,z,u)$ transform metaplectically under modular transformation by $S$ and $T$
(up to a phase) above eq. (39,50) and thus can be used to build cusp forms, as in the bosonic case, of the modular group
$A_l$ where $l=2|M^*/M|$.
$\lambda$ is any primary field. We have the ``Weyl--Kac"
formula for the ``characters",
$$\chi_\lambda(\tau,z,u)=A_{\lambda,k}/A_{0,0},\e$$
for $z=u=0$. These are not truly the characters of the RCFT,
except for $h=1$, but the characters can be presumably built in terms of these. The $\chi_{\lambda,k}(\tau,z,u)$ transform exactly by $S$ and $T$, eqs. (39,50), where $S$ has the sign $(-1)^r$ compared with eq. (39). To determine this sign, one need to analyze the modular transformations of $\chi_\lambda(p\tau)$ where $\chi_\lambda$ are the usual affine characters at $p=1$.

Now to prove the Gauss sum eq. (49) we calculate the central charge $c$ in two different ways. From the $S$ matrix eq. (39) we calculate $(ST)^3$ and get the gauss sum up to the phases defined in eq. (39). 
This gives the l.h.s. of eq. (51).
On the other hand we can compute it directly from the characters, eq. (55) which transform by this $S$. This gives the r.h.s. of eq. (51) finishing the proof.

Example: Take $G\approx SU(2)_{k/p}$. Then $h(\lambda)=p\lambda$. and the gauss sum is, eq. (49),
$$e^{\pi i \Theta/4}={1\over \sqrt{2(k+2)}}\sum_{m=0}^{2k+1}
e^{\pi i p m^2/(2(k+2))},\e$$
which is equal by eq. (52) to
$$\Theta=3p-2+4s+4r\mod 8,\e$$
where $s$ is defined by
$$px=1\mod 2(k+2),\qquad  1\leq (-1)^s x\leq k+1,\e$$
for some $x$.

Actually, we can use the results of the first part of this note to calculate the gauss sum for any lattice, not necessarily a group lattice, by writing
$$G\approx \osum_i SU(p_i^{a_i})\approx M^*/M,\e$$
We will assume that the automorphism $h$ is regular. Then it is enough to compute the gauss sum for $SU(N)_1$ and for any regular $h$.

To compute it for $SU(N)_1$ we note that by the Frenkel--Kac
construction \r\Kac, for any simply laced $G$,
$${A_{\lambda,g+1}^p(\tau,z,u)\over A^p_{0,g}}=
{\theta^M_{\lambda,1}(\tau,z,u)\over \eta(p\tau)^l}.\e$$
Thus, the central charges of $G^{\rm bosonic}_{1/p}$ and $G^{\rm affine}_{(g+1)/p}$ are the same. The latter is given by eq. (51) completing the calculation.

\appendix
 
In this appendix we will explicitly construct the lattices  realizing $SU(p^{m})_{1 \over q}$, where $p$ is a prime number, thus completing the realization for all the regular cases.
 
We use three types of matrices which depend on three integer parameters, $x,y,z$, obeying $x>0$, $y>0$ and $z^2<4 x y$. These matrices represent the scalar products between the basic vectors of the even lattices. 
The matrices are, of $E$ type,
$$E_{n}^{(x,y,z)} =\pmatrix{2 x &-z &0 &0  &\ldots&0&0&0&0\cr  
-z&2 y&-1 &0 &\ldots&0&0&0&0 \cr 
0&-1&2&-1&\ldots&0&0&0&0 \cr
0&0&-1&2&\ldots&0&0&0&0 \cr
\ldots&\ldots&\ldots&\ldots&\ldots&\ldots&\ldots&\ldots&\ldots& \cr
0&0&0&0&\ldots&2&-1&0&-1 \cr
0&0&0&0&\ldots&-1&2&-1&0 \cr
0&0&0&0&\ldots&0&-1&2&0 \cr
0&0&0&0&\ldots&-1&0&0&2},\e$$
for 
$$6 \le n \le 10,\qquad  \det(E_{n}^{x,y,z})=(4xy-z^{2})(11-n)-2x(12-n).\e$$

Of $A$ type,
$$A_n^{x,y,z}=\pmatrix{2 x &-z &0 &0  &\ldots&0&0&0&0\cr  -z&2 y&-1 &0 &\ldots&0&0&0&0 \cr 0&-1&2&-1&\ldots&0&0&0&0 \cr
0&0&-1&2&\ldots&0&0&0&0 \cr\ldots&\ldots&\ldots&\ldots&\ldots&\ldots&\ldots&\ldots&
\ldots& \cr
0&0&0&0&\ldots&2&-1&0&0 \cr
0&0&0&0&\ldots&-1&2&-1&0 \cr
0&0&0&0&\ldots&0&-1&2&-1 \cr0&0&0&0&\ldots&0&0&-1&2},\e)$$ 
$$ \det(A_n^{x,y,z})=(4 x y-z^2)(n-1)-2 x (n-2).\e$$
We also will use the one dimensional even lattice denoted by $SU(2)_{k}$, which is generated by the vector $\alpha^{2}=2 k$.

Our main findings are summarized in the tables. 
%
%
\topinsert

\line{\hfill Table 1: Realizations of $SU(2^m)_{1/q}$, where 
$m=2n+1$, $n>0$.\hfill}
\vskip15pt
\line{\hfill
\vbox{\offinterlineskip\hrule
\halign{&\vrule#&
  \strut\quad\hfil#\quad\hfil\cr
&q&&1&&-1&&3&&5&\cr
\noalign{\hrule}
&{\rm Central charge}&&7&&1&&9&&7&\cr
\noalign{\hrule}
&{\rm Realizations}&&$SU(2^{2 n+1})_{1}$&&$SU(2)_k, k=4^n$&&$E_9^{x,2,2},\ x={4+4^n \over 5}$&& $E_{7}^{x,1,1}$&\cr& && && &&$n$ even&& $x={2+4^n 
\over 3}$&\cr & && && &&$E_{9}^{x,2,1}, x={1+4^n\over 5},$&& &\cr& && && &&$n$ odd  && &\cr}
\hrule}\hfill}\vskip20pt\endinsert
%
%
%
%
%
%
\midinsert

\line{\hfill Table 2: Realizations of $SU(4)_{1/q}$.\hfill}
\vskip15pt
\line{\hfill
\vbox{\offinterlineskip\hrule
\halign{&\vrule#&
  \strut\quad\hfil#\quad\hfil\cr
&q&&1&&-1&&3&&5&\cr
\noalign{\hrule}
&Central charge  &&3&&5&&$9=1\mod8$&&7&\cr 
\noalign{\hrule} 
&Realizations&&$SU(4)_{1}$&&$D_5$&&$SU(2)_{2} \equiv D_9$  && $D_7$&\cr
}\hrule}
\hfill}
\vskip20pt\endinsert
%
%
%
\midinsert

\line{\hfill Table 3: Realizations of $SU(2^m)_{1/q}$, where $m=2n+2, n>0$.\hfill}
\vskip15pt
\line{\hfill
\vbox{\offinterlineskip\hrule
\halign{&\vrule#&
  \strut\quad\hfil#\quad\hfil\cr
&q&&1&&-1&&3&&5&\cr
\noalign{\hrule}
&Central charge&&7&&1&&5&&3&\cr 
\noalign{\hrule}
&Realizations&&$SU(2^{2 n+2})_{1}$&&$SU(2)_{k}, k=2^{n+1}$&&$A_5^{x,1,2}, x={8+2^{2 n+1}\over 5} ,$&&$A_3^{x,1,1}$&\cr
& && && &&n-even&& &\cr  
& && && &&$A_{5}^{x,1,1}, x={2+2^{2 n+1} \over 5},$&& $x={1+2^{2n+1} \over 3}$&\cr
& && && &&                         n-odd && &\cr}
\hrule}
\hfill}
\vskip20pt\endinsert
%
%
%
\topinsert

\line{\hfill Table 4: Realizations of $SU(p^m)_{1 \over q}$, where p-prime $p>2$, $p^{m} {\not =}1+8 k$.\hfill}
\vskip15pt
\line{\hfill
\vbox{\offinterlineskip\hrule
\halign{&\vrule#&
  \strut\quad\hfil#\quad\hfil\cr
&$p^{m}$&&Central charge&&Realizations&&Comments&\cr
\noalign{\hrule}
&$p^m=3+8 k, k \in Z$ &&  $c=6$       &&     $SU(p^m)_{-1} = E_6^{1+k,1,1}$           && &\cr 
&$p^{m}=7+8 k, k \in Z$&&$c=10$&&$SU(p^{m})_{-1} = E_{10}^{2(k+1),2,1}$&& &\cr
&$p^{m}=5+8 k, k \in Z$&&$c=8$&&$SU(p^{m})_{1 \over q} = E_{8}^{2(k+1),1,1}$&&See comment (1)&\cr
}\hrule}
\hfill}
\vskip20pt\endinsert

Comment (1):
Here $SU(p^{m})_{1}$ and $SU(p^{m})_{-1}$
are equivalent and both have the same central charge $c=4$. The new theory 
has central charge $c=8$.

Comment (2):
Here $SU(p)_1$ and $SU(p)_{-1}$ are equivalent and both have the same central charge $c=8$. The new theory has the central charge $c=4$.
%
\midinsert

\line{\hfill Table 5: 
Realizations of $SU(p^{m})_{1 \over q}$, where p-prime,  $p^{m}=1+8 k$.
\hfill}
\vskip15pt
\line{\hfill
\vbox{\offinterlineskip\hrule
\halign{&\vrule#&
\strut\quad\hfil#\quad\hfil\cr
&$N \equiv p^{m}$&&Central charge && Realizations &&  Comments&\cr
\noalign{\hrule}
&$N =p$-{\rm prime}, $N=24 k-7 $&&4&&$A_{4}^{1,2 k,1}$&& 
See comment (2) &\cr
&$N=p={\rm prime}$, $N=24 k+1 $&& 4 && $A_4^{x,y,z}$&& &\cr
& && &&$x,z$-odd,$y$-even&& &\cr
&$N$ is not prime&&8&&$A_{8}^{x,y,z},E_8^{x,y,z}$&& See examples below &\cr
}\hrule}
\hfill}
\vskip20pt\endinsert

Note that $|{M^{\star}/M}|={\rm det} M_{n}^{x,y,z}$, where $M_{n}^{x,y,z}$ stands for the lattices introduced above: $A_{n}^{x,y,z},E_{n}^{x,y,z}$. Therefore in order to prove that realizations for $SU(p^{m})_{1 \over q}$ are correct we have to check in most of the cases two conditions, namely: $\det(M_n^{x,y,z})=p^m$ and that there exist an element of the order $p^m$. The last condition ensures that the fusion rules are given by $Z_{p^m}$. The easiest way to check the second condition is to calculate one conformal dimension:
$$2 \Delta={{\rm det} M_{n-1}^{y,1,1} \over {{\rm det} M_{n}^{x,y,z}}}.\e$$
For example:
$$p^{m}=3+8 k, \hskip 0.3cm {\rm det} E_{6}^{1+k,1,1}=3+8 k, \hskip 0.3cm \Delta={2 \over p^{m}},$$ 
$$p^{m}=7+8 k, \hskip 0.3cm {\rm det} E_{10}^{2 (k+1),2,1}=7+8 k, \hskip 0.3cm \Delta={4 \over p^{m}},$$ 
$$p^{m}=5+8 k, \hskip 0.3cm {\rm det} E_{8}^{2 (k+1),1,1}=5+8 k, \hskip 0.3cm \Delta={1 \over p^{m}} \hskip 0.3cm etc.$$
The situation when $p^{m}=1+8 k$ is more tricky, here two different theories have the same central charge, but we found a lot of examples suggesting that the new theories may be realized by $A_{8}^{x,y,z},E_{8}^{x,y,z}$ lattices:
$$SU(9)_{-1} = A_{8}^{8,2,7},$$ 
$$SU(81)_{-1} = A_{8}^{2,2,1},$$
$$SU(25)_{1 \over 2} = E_{8}^{7,1,1},$$
$$SU(49)_{-1} = E_{8}^{13,1,1},$$
$$SU(121)_{-1} = A_{8}^{1,5,1}, $$
and so forth.

For the case of $N=24k+1={\rm prime}$ we have the examples,
$$SU(73)_{1\over5}=A_4^{5,2,3},$$
$$SU(97)_{1\over5}=A_4^{5,2,1},$$
$$SU(193)_{1\over5}=A_4^{11,2,3},$$
$$SU(241)_{1\over7}=A_4^{17,2,3},$$
$$SU(337)_{1\over5}=A_4^{17,2,1},$$
$$SU(457)_{1\over5}=A_4^{23,2,1},$$
and more. 
\refout\bye

\refout

\bye